\documentclass[useAMS,usenatbib]{mn2e}

\usepackage{graphicx}

\title[{\textit{UV}} Fe~{\sc{ii}} Emission in Quasars in Overdense Regions]
{Evidence of Increased \textit{UV} Fe~{\sc{ii}} Emission in Quasars in Candidate Overdense Regions}
\author[K.A.~Harris  et al.]{Kathryn A.~Harris$^{1}$\thanks{E-mail:
kateharris142@gmail.com}, G.M.~Williger$^{2,3,4}$, L.~Haberzettl$^2$,
 S.~Mitchell$^{2,5}$, 
 D.~Farrah$^1$, 
M.J.~Graham$^6$, R.~Dav{\'e}$^{7}$, M.P.~Younger$^8$, I.K.~S{\"o}chting$^9$\\ 
$^1$ Department of Physics, Virginia Tech, Blacksburg, VA 24061, USA \\
$^2$Department of Physics and Astronomy, University of Louisville, Louisville, KY 40292, USA \\ 
$^3$Lab. Lagrange, U. de Nice, UMR 7293, 06108 Nice Cedex 2, France\\ 
$^4$Institute for Astrophysics and Computational Sciences, Catholic University of America, Washington DC
20064, USA\\ 
$^5$Department of Aerospace Engineering ACCEND, University of Cincinnati,
Cincinnati OH, USA \\ 
$^6$Center for Advanced Computing Research, California Institute of
Technology, 1200 E California Blvd, Pasadena CA 91125, USA \\ 
$^{7}$Department of Astronomy, University of Arizona, 933 North Cherry Ave, Tucson, AZ
85721, USA\\
$^8$Jeremiah Horrocks Institute, University of Central Lancashire, Preston, PR1 2HE\\ 
$^9$Astrophysics, Denys Wilkinson Building, Keble Road, University of Oxford, Oxford OX1 3RH \\
}

\date{Accepted xxxx. Received xxxx; in original form xxxx}

\begin{document}


\maketitle


\begin{abstract}
We present evidence for a skewed distribution of \textit{UV} Fe~\textsc{ii} emission in 
quasars within candidate overdense regions spanning spatial scales of $\sim50$ Mpc at 
$1.11 < z < 1.67$, compared to quasars in field environments at comparable redshifts. The 
overdense regions have an excess of high equivalent width sources (W2400 $>$ 42 \AA), 
and a dearth of low equivalent width sources. There are various possible explanations for 
this effect, including dust, Ly$\alpha$ fluorescence, microturbulence, and iron abundance. 
We find that the most plausible of these is enhanced iron abundance in the overdense regions, 
consistent with an enhanced star formation rate in the overdense regions compared to the field. 

\end{abstract}


\section{INTRODUCTION}
There is significant 
controversy over the stellar mass-metallicity (M-Z) relation as a function of environment 
and redshift. The general expectation might be that metallicity is higher in overdense 
regions at a given redshift, since high redshift starburst galaxies seem to prefer 
such regions \citep{Gomez2003,Blain2004,Farrah2006,Cooper2008}. Earlier star formation 
would give rise to earlier metal enrichment of the ISM/IGM. For example, supernovae 
\citep[e.g.][]{Adelberger2005,Domainko2004} may efficiently enrich the IGM over Mpc scales.

Conversely, direct observational studies are ambiguous. At low redshift, some authors 
\citep[e.g.][]{Hughes2013} find no relation between metallicity and environment, 
while others \citep{Skillman1996,Cooper2008} claim a weak but 
significant trend for galaxies in groups or clusters to have higher metallicities than 
field galaxies. 
At higher redshifts, there is even more uncertainty \citep[e.g.][]{Hamann1993} with few 
studies considering environment.

A potentially powerful way to constrain star formation histories in different environments 
at high redshifts is to use the ratio of Fe ~\textsc{ii}[\textit{UV}] to Mg~\textsc{ii}[$\lambda$2798]. 
To first order, Fe ~\textsc{ii} is produced from SNeIa roughly one Gyr after the initial burst 
of star formation, while Mg~\textsc{ii} is created in SNeII. Hence their ratio can be used 
as a cosmological clock \citep{Hamann1993} to age-date the initial star formation. 
Moreover, both emission lines are seen in quasars, where the quasar illuminates 
the metal rich gas. This allows the lines and therefore the metallicities to be observed 
to potentially very high redshifts. 
However there is a large amount of scatter seen in this ratio, the reasons for which 
are not fully understood.

\textit{UV} Fe~\textsc{ii}
has been observed in different objects such as symbiotic stars \citep[e.g][]{Hartman2000},
young stellar objects \citep[e.g.][]{Cooper2013}, novae \citep[e.g][]{Johansson1984} and 
the Broad Line Region (BLR) of active
galactic nuclei (AGN) \citep{Sigut1998}.
In AGN, \textit{UV} Fe~\textsc{ii} is seen
at varying strengths, though the reasons for this variation are still debated.
A number of Fe~\textsc{ii}-bright quasars have been found and studied in detail over a wide 
redshift range \citep[e.g.][]{Osterbrock1976,Weymann1991,Graham1996,Vestergaard2001,Bruhweiler2008}.

While several mechanisms likely affect the observed iron emission 
(e.g. abundance, collisional excitation, microturbulence and Ly$\alpha$ fluorescence, see e.g. 
\citealt{Netzer1983,Sigut2003, Baldwin2004,Matsuoka2007}), it is plausible (given that all but abundance 
are small $<$pc scale mechanisms and unlikely to be effected by the $>$Mpc scale environment)
that this emission is a reasonably proxy for the metallicity build up in galaxies.

In this paper, we explore the use of the \textit{UV} Fe~\textsc{ii} in high redshift quasar spectra to consider 
differences in SFHs in different environments at high redshift. To do so, we consider 
the overdense regions of quasars in Large Quasar Groups (LQGs). 

LQGs are some of the largest candidate overdensities seen in the Universe, spanning 50-200 
$h^{-1}$ Mpc, have been found at $z > 1$, and are potentially the precursors of the large 
overdensities seen at the present epoch, such as super-clusters \citep*{Komberg1996,Wray2006}. 
These LQGs exist at high redshifts and presumably trace the mass distribution.  
There are $\sim$ 40 published examples of LQGs \citep[][(CCGS12) and references
therein]{Clowes2012}.

By using LQGs we can quickly assemble statistically significant numbers of quasars in overdense 
regions, to compare to field samples. 
The observations for this paper were taken in the direction of the
Clowes-Campusano LQG (CCLQG; \citealt{Clowes1991,Clowes1994}) which lies at a redshift of
$z \sim 1.3$, and spans $\sim$100-200  $h^{-1}$ Mpc.

We compare the \textit{UV} Fe~\textsc{ii} in quasars in LQGs at $z > 1$ to the same emission seen in quasars in 
the field over similar redshifts to search for differences in star formation history 
as a function of environment. 
We will present 12 AGN at $z$ = [1.159,1.689] with increased
\textit{UV} Fe~\textsc{ii} emission (W2400 $>$ 32\AA) evident in their spectra. All of the
quasars are within an area of 1.6 deg$^{2}$, and lie within the redshift range of the 
overdensity previously described.
The cosmology used is $H_0=70$ kms$^{-1}$Mpc$^{-1}$, $\Omega_m=0.27$ and
$\Omega_{\Lambda}=0.73$.

\section{ANALYSIS}

We treat the LQG region as a potential high density environment.

By comparing the measurements of the Fe~\textsc{ii} emission in these quasars to the 
emission from a control sample of randomly selected quasars, we examine any 
differences between the 
samples. Due to the limits of the observations, we 
do not study the whole LQGs field, using only two 0.8 deg$^2$ of the area (which is  
covered by our additional observations described later in this section). These fields are 
centred on RA = 162.146, Dec = 5.406, and RA = 162.514, Dec = 4.528.

\subsection{{FE~{\sc{ii}}} MEASUREMENT TECHNIQUES}\label{sect:spectra}
To measure the Fe \textsc{ii} emission, we used the method described in 
\citet{Weymann1991}. We use this method to provide an estimate of the
overall emission as opposed to, for example, the \citet{Hartig1986} method which gives an
estimate at a single wavelength. The
continuum level is found at the central wavelength within two wavelength ranges,
2240--2255 {\AA} and 2665--2695 {\AA}. A straight line is then drawn between the centres
of these two wavelength ranges to create the effective continuum. 
\citet{Weymann1991} calculate the equivalent width (EW)
between 2255 and 2650 {\AA} (W2400) with respect to this effective continuum level.
The errors on the measurements are estimated based on the noise across the continuum which 
has the greatest effect and therefore the dominant error 
in the Fe~\textsc{ii} measurement. (The values are estimated in 
Section \ref{sect:control}.)

\begin{figure*} \begin{center} \begin{tabular}{cc}

\includegraphics[width=0.45\textwidth]{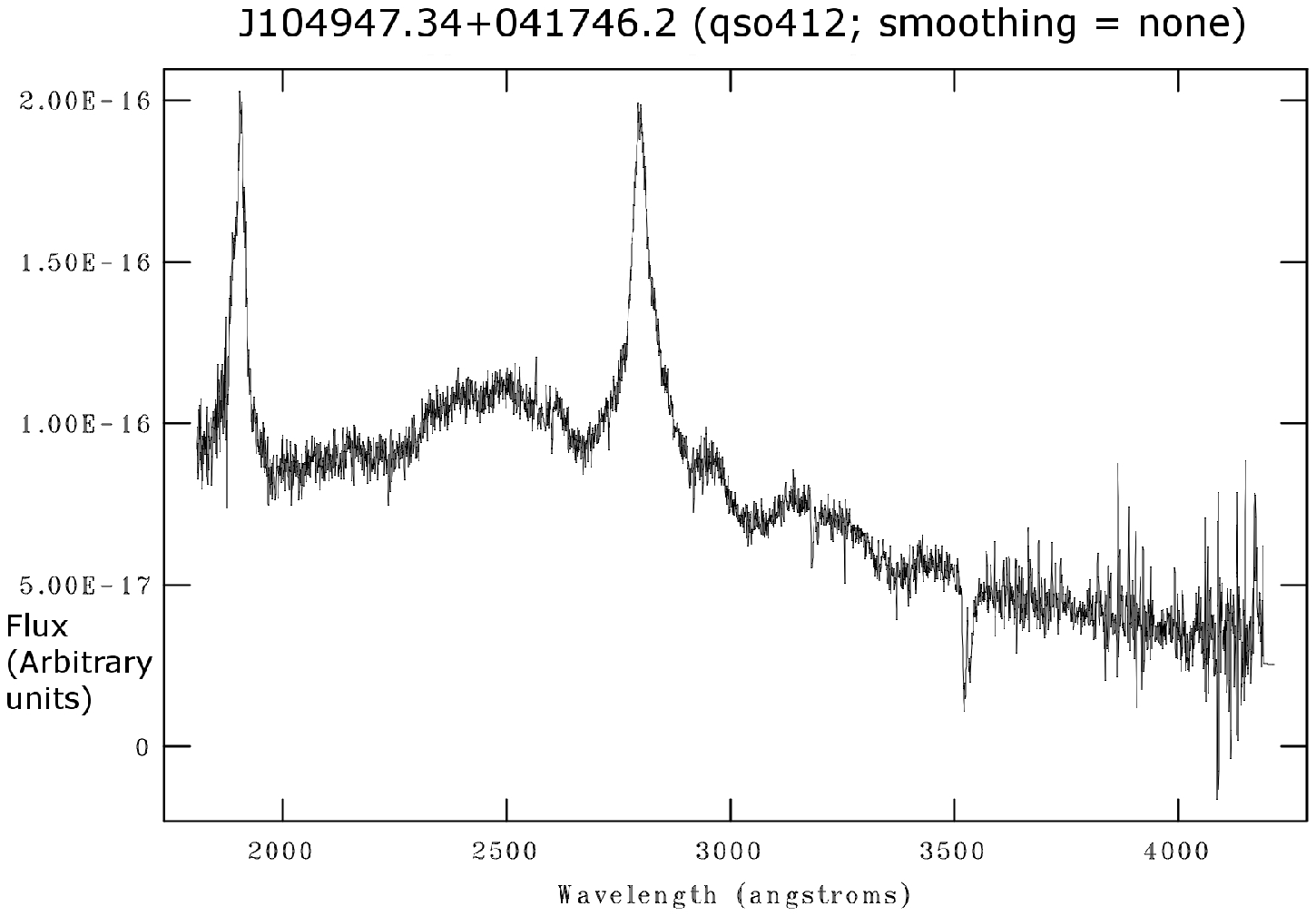}  &
\includegraphics[width=0.45\textwidth]{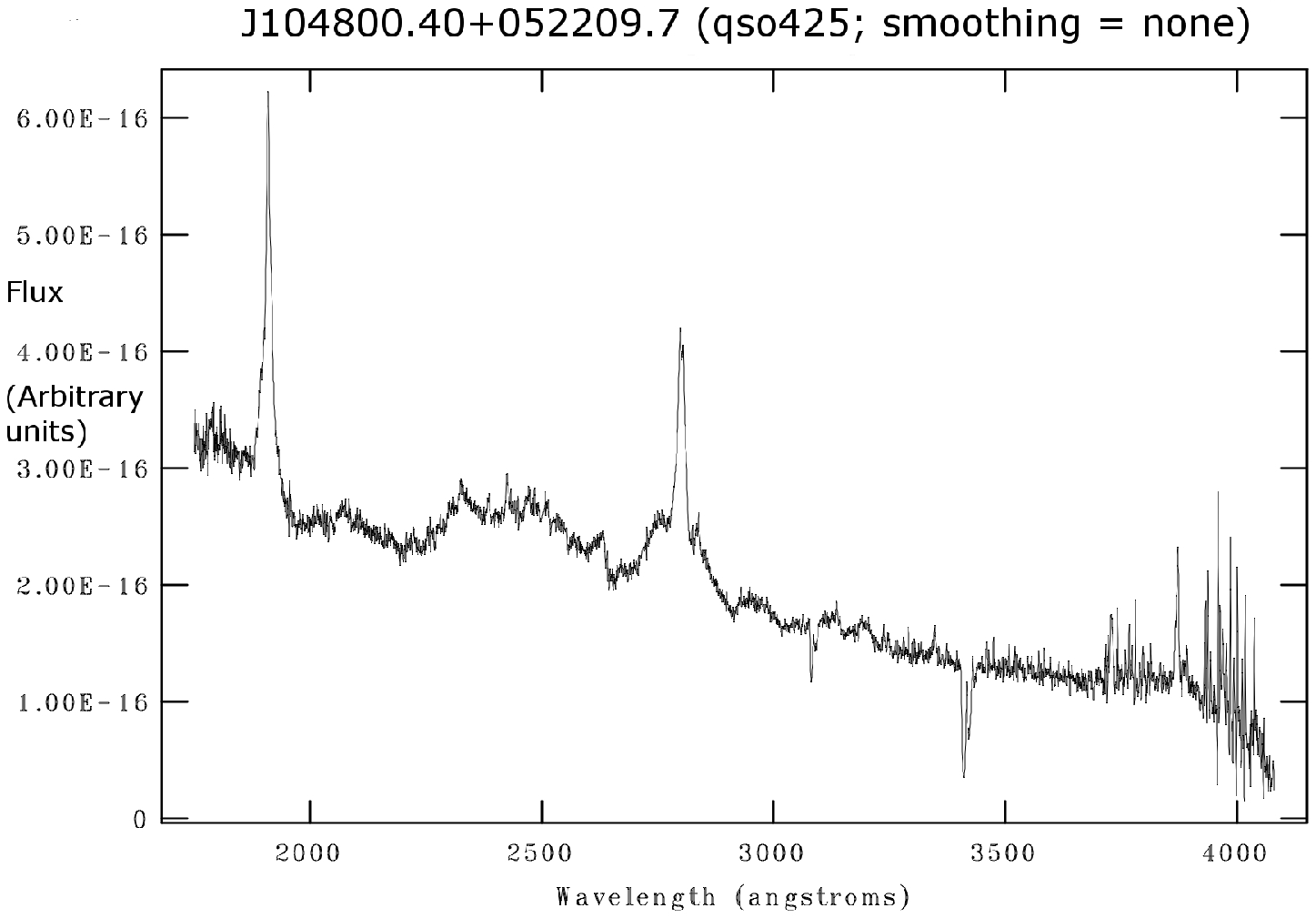}  \\
\includegraphics[width=0.45\textwidth]{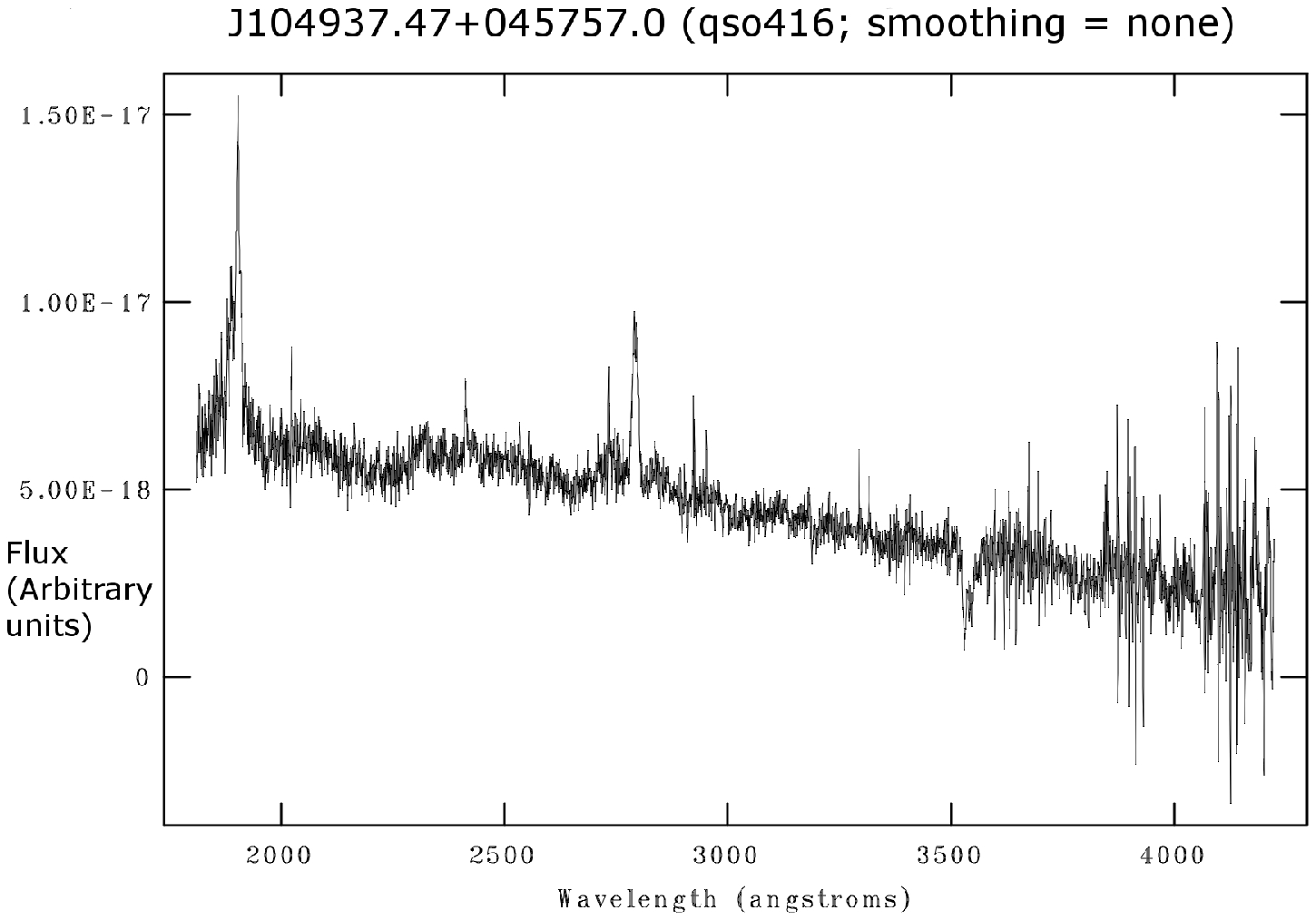} &
\includegraphics[width=0.45\textwidth]{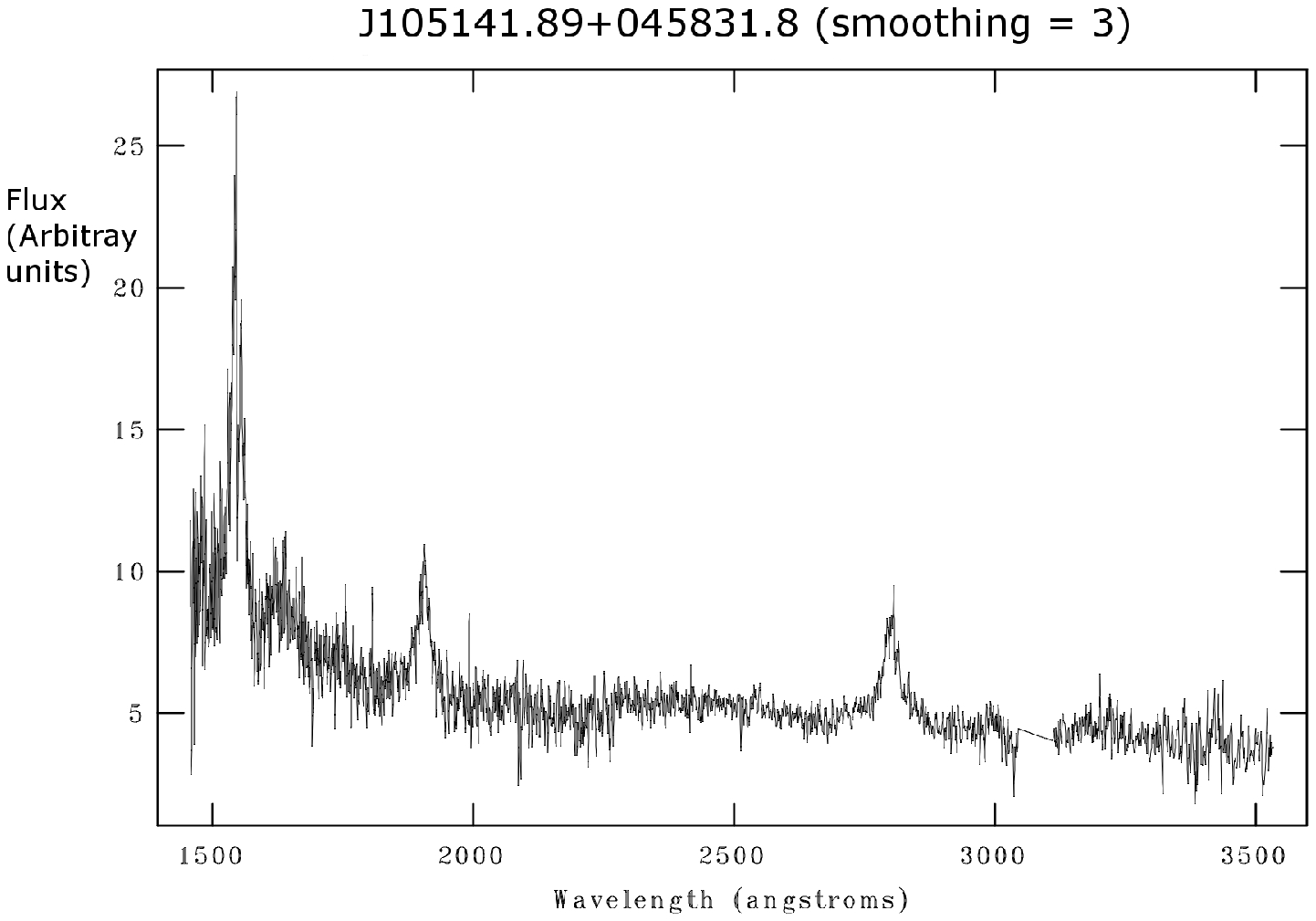} \\ 
\end{tabular}
\end{center} \caption{\small{Example LQG spectra with a range of emission strengths 
from Hectopsec (J104947.34+041746.2 (qso412),
J104800.40+052209.7 (qso425), and J104937.47+045757.0 (qso416) shown) and SDSS (J105141.89+045831.8 
shown). All spectra can be found online.}} 
\label{fig:spec}
\end{figure*}

\subsection{LQG FIELD SAMPLE}
Two LQGs and an additional quasar set have been found in the area studied in this paper. 
The overdensity was estimated using $(\rho - \langle\rho\rangle)/\langle\rho\rangle$ (CCGS12).
\begin{enumerate} 
\item L1.28: The CCLQG lies at $z$ = [1.187,1.423], contains 34 members,
and has an estimated overdensity of 0.83 and a statistical significance of 3.57$\sigma$
(CCGS12). 
\item L1.11: There is another LQG at $z$ = [1.004,1.201], containing 38 members (CCGS12). 
This group has an estimated overdensity of 0.55
and a statistical significance of 2.95$\sigma$. 
\item L1.54: There is an ``enhanced set''
of quasars with 21 members at $z$ = [1.477,1.614]. This group has an overdensity of 0.49, and a
statistical significance of 1.75$\sigma$, which though suggestive, is not high enough to
be statistically significant for a large structure (\citealt{Newman1999}, CCGS12).
\end{enumerate}

The original LQG members were selected from the SDSS DR7QSO catalogue
\citep{Schneider2010}. A magnitude cut of $i$-mag = 19.1 \citep{Schneider2010} was applied to create a uniform
sample and quasars are within a 3D linking length of 100 Mpc. A convex hull is created
around the members, giving the total volume covered by the LQG. See CCGS12 for more details 
on the method used to select LQG members.

The latest discussion of these LQGs can be found in CCGS12. Due to uncertainties over LQG
membership caused by the member selection criteria and sample completeness, for the rest of the paper we will not be discussing
the LQG or which quasars are classed as members. We will assume that quasars trace the
mass distribution and therefore this area space and redshift range is therefore a candidate
overdense region.
\citet[][and references therein]{Martini2013} found 
for $1 < z < 1.5$ the fraction of AGN lying in clusters is increased compared to lower redshifts, 
making this a reasonable assumption.

There are 10 quasars at $1.1 < z < 1.7$ from the SDSS DR7 QSO catalogue
\citep{Schneider2010} which have SDSS spectra in the area of the LQGs we are studying.  
 The spectra cover the wavelength range 3800--9200 {\AA} and have
a resolution of 2.5 {\AA} \citep{SDSSprojectbook}.

To improve statistics and better sample the overdensity, we increased the sample size.
We start with a sample of quasars with photometric 
redshifts from the DR7 catalogue by \citet{Richards2009} which place them within the 
redshift range of the LQGs. We then randomly selected 
a subset of 32 for followup spectroscopy (observed as part of a larger observing project), dependent on available 
fiber positioning. We used the Hectospec instrument \citep{Fabricant2005} 
a multi-object optical spectrograph, mounted at the 6.5-m MMT on Mount
Hopkins, Arizona. The spectra were taken over nine nights and, due to inaccuracies in 
photometric redshifts, produced 18 quasar spectra within the required redshift range.  
The remaining objects were a mixture of quasars (generally at lower redshifts) and star forming galaxies.

The Hectospec data cover 3900 to 9100 {\AA} and have a resolution of 1.2 {\AA}. These
spectra were reduced using the IDL based pipeline, \textsc{HSRED}\footnote{HSRED is an IDL based
reduction package for Hectospec spectra created by Richard Cool and hosted at
http://www.astro.princeton.edu/$\sim$rcool/hsred/}. Table \ref{tab:feIIobslog} shows 
the dates, fields, and exposures times for the Hectospec observations.

\begin{table} 
\caption{\small{Observing log for the Hectospec data.}}
\begin{tabular}{@{}cccc@{}} 
\hline 
Date     & RA (J2000)   & Dec (J2000) & Exposure (s) \\ \hline 
17.02.2010  & 10:50:16.9 & +04:37:12 & 5400   \\ 
18.02.2010  & 10.50:16.9 & +04:37:12 & 5400  \\ 
19.02.2010  & 10:50:06.9 & +04:29:16 & 5094  \\ 
06.04.2010  & 10:50:06.9 & +04:29:16 & 5400  \\ 
07.04.2010  & 10:48:31.8 & +05:23:29 & 7200  \\
09.04.2010  & 10:48:31.8 & +05:23:29 & 5400  \\ 
10.04.2010  & 10:48:38.9 & +05:25:57 & 5400  \\ 
11.04.2010  & 10:48:38.9 & +05:25:57 & 5400  \\ 
11.04.2010  & 10:49:57.0 & +04:30:01 & 5400  \\ 
12.04.2010  & 10:49:57.0 & +04:30:01 & 1800  \\ \hline 
\end{tabular}
\label{tab:feIIobslog} 
\end{table}

The final catalogue of quasars (see Table \ref{tab:Fedata}) contains 18 quasars from 
Hectospec and 6 quasars from SDSS spectra 
within the redshift range $1.1 < z < 1.7$. Four of the SDSS quasars were removed due to 
low signal-to-noise spectra but are included in Table \ref{tab:Fedata} for completeness. 
The area occupied by these quasars covers 1.6 deg$^2$ of the LQGs. An example of the spectra 
is shown in Fig. \ref{fig:spec}.

\subsubsection{COMPLETENESS AND LQG MEMBERS}\label{sect:complete}

The Hectospec quasars, though not a complete sample, were randomly selected across area 
and redshift range, with no bias towards strong or weak Fe~\textsc{ii} emission, magnitude, 
or location (beyond being within the field of the LQG overdensities). The quasars were observed 
as part a larger project which observed lower redshift luminous red galaxies. Therefore there 
was no biasing on the placement of the available fibers for observing these quasars. 

Because of the data and the above described member selection method, we can say which  
quasars are part of the LQG as it is defined in CCGS12 but can not say whether these are 
the only members. 
If the sample used to determine members were complete down to the magnitude of 
$g$-mag = 21.1 (limit of the Hectospec data), additional members 
may be found and the shape of the convex hull would change.

For the purposes of this paper, we will assume that the LQGs indicate a general 
overdensity within this region.
When mentioning the LQGs region, we refer to a region of space with a potential mass 
overdensity.

\subsection{CONTROL SAMPLE}\label{sect:control}

The control sample was taken from
Stripe 82 from SDSS \citep{York2000}, which has a similar limiting magnitude 
(complete down to \textit{g}-mag=21 to match the general completeness in the area of the 
LQGs) and taken from areas which do not contain any previously known LQGs.
The samples were run through the program used to find the LQG 
and were determined not to be within a LQG within a 2$\sigma$ significance. 
We took multiple two deg$^{2}$ samples across 
the length of the stripe to reduce the impact of any marginal overdensities in a single area.
The initial sample contains in total 394 field quasars within the redshift range 
$1.1 < z < 1.7$.

The errors were estimated across a range of objects and compared to the measured SNR. 
Spectra with SNR$ < 5$ per pixel rest EW had
errors of $\pm$ 8.4 \AA. This decreases to $\pm$ 4.8 \AA\ for $5 <$ SNR $< 10$ per pixel
and $\pm$ 2.85 \AA\ for SNR $> 10$ per pixel.
Therefore, to reduce the effects of errors in measurements due to low SNR, spectra with an 
average SNR $<$ 5 per pixel were rejected. This removes four quasars from the LQGs field 
leaving 24, and reduces the control sample to 178 quasars, removing 
more control quasars due to generally lower SNR in SDSS spectra. The rejected quasars cover 
a range of W2400 EW values and do not favour any strength.

\begin{figure} 
\includegraphics[width=60mm,angle=-90]{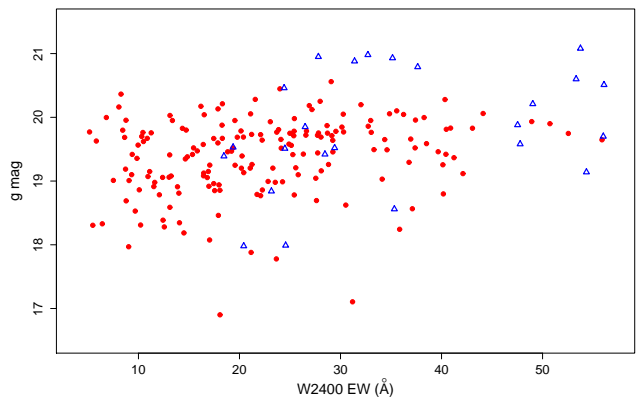}
\caption{\small{Comparison of the distribution of W2400 EW (\AA) as a function of
\textit{g} magnitude for the quasars in the LQGs field (triangles, blue online) and
control samples (circles, red online).}} 
\label{fig:W2400_mag} 
\end{figure}

Fig. \ref{fig:W2400_mag} shows the distribution of W2400 EW as a function of the $g$ band 
magnitude for both the control sample (circles, red) and the LQG field quasars (triangles, blue). Though some of 
the Hectospec quasars are fainter than the control sample quasars, there is no obvious 
relation between the magnitude and the W2400 EW emission. This is discussed further in 
section \ref{sect:discuss}.

\begin{table*} 
\caption{\small{Properties of the Hectospec quasars along with the
properties for any other quasars within the 1.6 deg$^2$ field from the SDSS DR7QSO
catalogue \citep{Schneider2010}. The columns show the Fe~\textsc{ii} group,
names, RA, DEC, redshift, LQG membership, Fe~\textsc{ii} measurements using the method
described in \citet{Weymann1991}, and the $g$-magnitude, taken from SDSS. The asterisk on
the quasar ID indicates previously known LQG members. L1.11 denotes the group with $\overline{z} = 1.11$ (CCGS12),
L1.28 for the group with $\overline{z} = 1.28$ (CCGS12) and L1.54 indicates the additional quasar set at $\overline{z} = 1.54$ 
(unpublished data). }}
{\small \begin{tabular}{c l | c c c c c c }
\hline 
Group          & Quasar                                           &  Redshift & RA (J2000) & DEC (J2000) &  membership$^a$ & W2400$^b$ (\AA) & \textit{g}-mag \\ \hline 
Ultra-strong   & SDSS J104947.34+041746.2/qso412     			   & 1.159  & 10:49:47.35 & +04:17:46.35 & L1.11    & 56.08         & 20.51      \\    
               & SDSS J104800.40+052209.7/qso425      			   & 1.230  & 10:48:00.41 & +05:22:09.90 & L1.28    & 56.01         & 19.70      \\ 
               & SDSS J104914.32+041428.6*                         & 1.607  & 10:49:14.33 & +04:14:28.65 & L1.54    & 54.34         & 19.14     \\ 
               & SDSS J104930.44+054046.1/qso27                    & 1.315  & 10:49:30.46 & +05:40:46.20 & L1.28    & 53.75         & 21.08       \\ 
               & SDSS J104815.93+055007.8/qso421                   & 1.665  & 10:48:15.94 & +05:50:07.80 &          & 53.32         & 20.60     \\ 
               & SDSS J104926.83+042334.6/qso417                   & 1.653  & 10:49:26.83 & +04:23:34.80 &          & 49.03         & 20.21     \\ 
               & SDSS J104921.05+050948.3/qso29                    & 1.417  & 10:49:21.07 & +05:09:48.30 &          & 47.78         & 19.58     \\ 
               & SDSS J105131.95+045124.7/qso41                    & 1.434  & 10:51:31.94 & +04:51:24.90 &          & 47.53         & 19.88     \\ \hline
Strong         & SDSS J104958.91+042723.3/qso217                   & 1.622  & 10:49:58.92 & +04:27:23.40 & L1.54    & 37.64         & 20.79      \\ 
               & SDSS J105010.05+043249.1/qso48*                   & 1.217  & 10:50:10.06 & +04:32:49.20 & L1.28    & 35.33         & 18.56      \\ 
               & SDSS J104933.41+054840.3/qso219                   & 1.349  & 10:49:34.71 & +05:48:36.00 & L1.28    & 35.15         & 20.93     \\ 
               & SDSS J105255.65+055112.9$^{c}$                    & 1.678  & 10:52:55.65 & +05:51:12.93 &          & 32.8          & 20.03      \\ 
               & SDSS J104937.47+045757.0/qso416                   & 1.154  & 10:49:37.48 & +04:57:57.10 &          & 32.72         & 20.98     \\ \hline 
Average        & SDSS J105000.36+045157.8/qso410                   & 1.418  & 10:50:00.36 & +04:51:57.90 &          & 31.39         & 20.88     \\ 
               & SDSS J105154.14+041059.5$^{c}$                    & 1.552  & 10:51:54.14 & +04:10:59.55 & L1.54    & 29.94       & 21.29       \\ 
               & SDSS J105141.89+045831.8*                         & 1.608  & 10:51:41.91 & +04:58:27.90 & L1.54    & 29.42         & 19.52   \\ 
               & SDSS J105007.90+043659.7/qso49                    & 1.131  & 10:50:07.90 & +04:36:59.70 & L1.11    & 28.46         & 19.42        \\ 
               & SDSS J105036.09+045608.3/qso45                    & 1.317  & 10:50:36.10 & +04:56:11.40 & L1.28    & 27.81         & 20.95        \\ 
               & SDSS J105352.75+043055.0/qso22                    & 1.216  & 10:50:30.77 & +04:30:55.05 & L1.28    & 26.5          & 19.85         \\ 
               & SDSS J104656.71+054150.3*                         & 1.228  & 10:46:56.71 & +05:41:50.25 & L1.28    & 24.57         & 17.99      \\ 
               & SDSS J104751.88+043709.9                          & 1.696  & 10:47:51.89 & +04:37:09.90 &          & 24.49         & 19.51      \\ 
               & SDSS J104840.85+040938.3/qso420                   & 1.238  & 10:48:40.85 & +04:09:38.55 &          & 24.42         & 20.46     \\
               & SDSS J105249.68+040046.3$^{c}$                    & 1.193  & 10:52:49.68 & +04:00:46.50 & L1.11    & 24.12      & 19.27      \\ 
               & SDSS J104932.22+050531.7/qso26*                   & 1.111  & 10:49:32.23 & +05:05:31.50 & L1.11    & 23.16         & 18.84         \\ 
               & SDSS J104733.16+052454.9*                         & 1.334  & 10:47:33.17 & +05:24:55.05 & L1.28    & 20.42         & 17.98          \\ 
               & SDSS J104943.28+044948.8/qso413                   & 1.295  & 10:49:43.30 & +04:49:48.75 & L1.28    & 19.37         & 19.53         \\ 
               & SDSS J104938.35+052932.0*$^{c}$                   & 1.517  & 10:49:38.35 & +05:29:31.95 & L1.54  & 18.83     & 19.48       \\ 
               & SDSS J105018.10+052826.4*                         & 1.307  & 10:50:18.12 & +05:28:26.40 & L1.28    & 18.46         & 19.39    \\   \hline
\end{tabular}
}
\label{tab:Fedata} \\ 
\begin{flushleft} \scriptsize{$a.$ The membership is decided by quasar redshift and its inclusion within a
convex hull created from the list of previously known members.} \\ 
\scriptsize{$b.$ Though the values can not be measured to this number of significant
figures due to errors, the data has been left at two decimal places in order to remove the
problems of ties in the data when running the Mann-Whitney test (described further in
Section \ref{sect:w2400dist}).} \\ 
\scriptsize{$c.$ These quasars are within the area of
the LQGs and additional candidate overdensity. However, they will not be included in the statistics due to low
signal-to-noise in the spectra.} \\ 
\end{flushleft}
\end{table*}

\begin{figure} \begin{center} 
\includegraphics[width=0.35\textwidth,angle=-90]{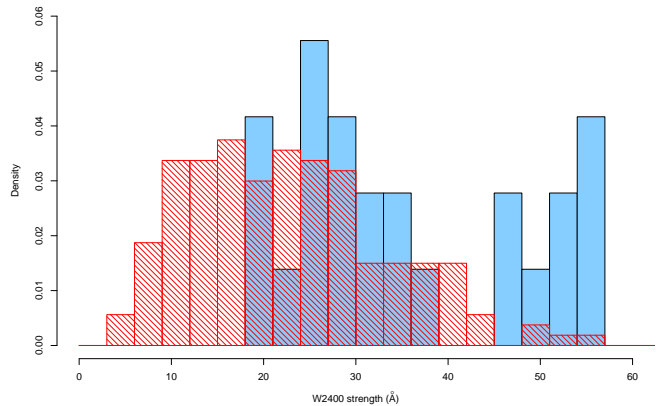} 
\end{center} 
\caption{\small{Normalised histogram (bin size= 3{\AA}) showing the distribution of the measured W2400 EW
density for the LQGs field (solid, blue online, 24 objects) and the control sample
(hatched, red online, 178 objects).}} 
\label{fig:hist_dist_both} 
\end{figure}

\section{RESULTS}

Table \ref{tab:Fedata} summarises the data for the sample. The quasars with an SDSS name as well 
as qsoXXX are those quasars selected from the photometric catalogue and re-observed 
using Hectospec. The spectra for these objects is available in the online material for this paper.
The four quasars 
removed from the LQG sample due to low SNR have been included for completeness (denoted by ``c'') but are 
not included in the analysis.

Table \ref{tab:sample_props} shows the median, mean and standard 
deviation of the control and LQGs samples. These data were used to define boundaries;
the representative average range for the
Fe~\textsc{ii} equivalent width was taken as 10 - 32 \AA, anything 
between 32 and 43 \AA\ EW was classed as strong and greater than 43 \AA\ EW
was classed as ultra-strong Fe~\textsc{ii}.

Using this system, eight quasars were classed as ultra-strong and four were classed as strong Fe~\textsc{ii}
emitters from a sample size of 24 quasars within 1.6
deg$^{2}$, in the redshift interval of $1.1 < z < 1.7$.

\subsection{A SIGNIFICANT DIFFERENCE IN THE DISTRIBUTION OF ULTRA-STRONG EMITTERS} 

Table \ref{tab:cont_samples} shows the number of quasars
(and percentage) with different \textit{UV} Fe~\textsc{ii} strengths in the LQGs field and the
control fields. We show both the complete sample and a magnitude limited sample where all the 
quasars are within the same magnitude range ($17.98 < g < 20.56$).
The LQGs field has a large percentage of quasars with 
strong and ultra-strong Fe~\textsc{ii} emission.
33.3 $\pm~11.8$ per cent of the LQG field sample show ultra-strong Fe~\textsc{ii} 
emission and 16.7 $\pm~8.3$ per cent show strong emission. This compares to the control sample 
which has 3.4 $\pm~1.4$ per cent of quasars showing ultra-strong emission and 15.7 $\pm~3.1$ per cent
showing strong emission. 
Thus there is a statistical difference for the ultra-strong emitting quasars, which is also seen 
to the magnitude limited samples. 
For the magnitude limited samples, the percentage of strong quasars in the LQG field drops 
to 5.9 $\pm~5.9$ per cent, compared to the control sample value of 16.0 $\pm~3.0$ per cent, 
which are no longer within the errors.

However as the definitions of strong and ultra-strong are arbitrary and dependent on the 
control sample, for the rest of the paper, we will concentrate on the differences in the 
full distribution from the data and control samples. 

\begin{table} 
\caption{\small{Median, mean and standard deviation of W2400 EW (\AA) for the 
LQG sample and SDSS control sample.}}
\centering 
\begin{tabular}{c|ccc} 
\hline 
Sample          & median    & mean    & standard \\
                &           &         & deviation           \\  \hline 
control sample  & 21.20     & 22.59   & 10.86               \\ 
LQG field       & 32.05     & 35.71   & 12.71               \\ \hline
\label{tab:sample_props} 
\end{tabular} 
\end{table}

\begin{table*} 
\caption{\small{The number of quasars per deg$^2$ for the different
Fe~\textsc{ii} strengths for quasars in the control fields (covering a total area of 26 deg$^2$) 
compared to the LQGs field (which cover 1.6 deg$^2$). The percentages are those of 
the total in the field. The magnitude limited sample has the magnitude range $17.98 < g < 20.56$. }} 
\centering 
\begin{tabular}{c|cc|cc} 
\hline 
             & \multicolumn{2}{|c|}{Complete}   & \multicolumn{2}{|c}{Mag. Limited}  \\ 
Strength     & LQGs field      & control field  & LQGs field      & control field   \\  \hline 
Ultra-strong & 5.0  (33.3\%)   & 0.23 (3.37\%)  & 3.75  (35.3\%)  & 0.23 (3.45\%)   \\ 
Strong       & 2.5  (16.7\%)   & 1.08 (15.7\%)  & 0.63  (5.9\%)   & 1.08 (16.2\%)  \\
Average      & 7.5  (50.0\%)   & 4.78 (69.7\%)  & 6.25  (58.8\%)  & 4.62 (69.4\%)  \\
Weak         & 0.0  (0.0\%)    & 0.77 (11.2\%)  & 0.0   (0.0\%)   & 0.73 (11.0\%)  \\  \hline 
\label{tab:cont_samples} 
\end{tabular} 
\end{table*}

\begin{figure*} 
\includegraphics[width=120mm,angle=-90]{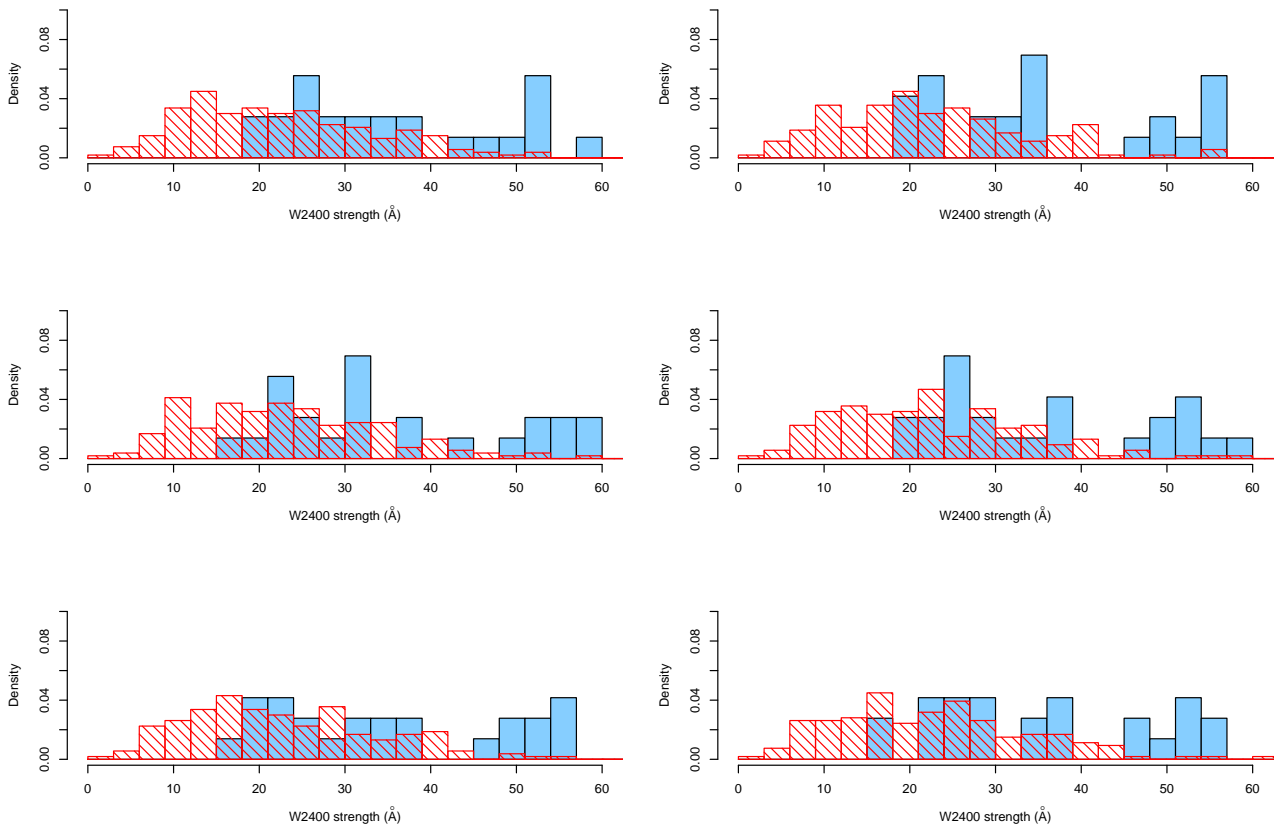}
\caption{\small{Example histograms (bin size = 3\AA) showing six possible distributions
within the errors of the measurements taken from the Monte-Carlo re-sampling. The
histograms shows the distribution of the measured W2400 EW for the LQGs field (solid, blue
online, 24 objects) and the control sample (hatched, red online, 178 objects). }}
\label{fig:hist_err} 
\end{figure*}

\subsection{W2400 distribution}\label{sect:w2400dist}

Fig. ~\ref{fig:hist_dist_both} (bin size= 3 {\AA}) shows the distribution of the W2400 EW for the LQGs field (solid,
blue online) and the control sample (hatched, red online). The relative excess of \textit{UV} Fe~\textsc{ii}
emission in the LQGs field can be clearly seen for W2400 EW $> 45$ \AA. For W2400 EW $< 20$ \AA, the
histogram shows the lack of low emission quasars within the LQGs field compared to the
control sample. 
Fig. \ref{fig:hist_err} shows a selection of histograms from a Monte-Carlo method. For
each histogram, a point is randomly selected for each object across the whole distribution with 
appropriate weighting. This figure shows at the upper end of the
emission, there is again an excess of quasars with W2400 $>$ 45, indicating this
result is not affected by the errors. There is also a lack at W2400 $<$ 20 \AA.

To quantify the difference in distributions, we employ the Mann-Whitney test, a
powerful non-parametric test for comparing two populations. The Mann-Whitney
test does not require any assumptions about the forms of the distributions, and is less
likely to apply significance to outliers due to the ranking method used. This test is however
sensitive to rounding, which can create ties in ranks in the data, therefore we have measured to two decimal places, 
though the data is not accurate to this level, and do not
rounded our data at any point \citep{DeGroot1986}. 
Median latencies in the LQG field and control sample are 32.05 and 21.20 respectively. 
Using a one-tailed Mann-Whitney test, with 24 LQG quasars and 178 control sample quasars, the 
distributions in the two groups differ significantly with a p-value of 99.996\%.

To estimate the effects of the errors on the W2400
measurements, a Monte-Carlo method was used to resample points from
within the error limits for each measurement across the whole distribution with appropriate weighting  
and Mann-Whitney test repeated, using the same parameters as above. 
In each case, P $<$ 0.05. 
Therefore taking into account errors, the two distributions are still differ significantly.

To investigate the lack of weak Fe~\textsc{ii} emitting (W2400 $<$ 20 \AA) which could be due to the limit sample size, 
the Mann-Whitney statistical test was repeated using the samples with only W2400 $>$ 20 \AA. 
This test gives P = $0.013 \pm ~0.05$,
indicating that removing the weak emitters does have a significant effect on the result. 
However, this artificially truncates the values, creating an artificial distribution. 
To properly test this lack of weak emitters, a larger sample of quasars 
within overdense regions would be needed.

\section{DISCUSSION}\label{sect:discuss}

We have shown there is an increase in the Fe~\textsc{ii} emission within quasars within the LQGs 
compared to our control sample. There are various possible explanations:
\begin{enumerate}
\item a selection effect - created by the selection of LQG quasars and magnitude limits,
\item dust - different amounts of dust within the LQG sample and the control sample causing the 
difference in the observed EW distributions,
\item Ly$\alpha$ fluorescence - Ly$\alpha$ pumping can cause an increase the Fe~\textsc{ii}, 
\item microturbulence - motions within the cloud line emitting region, 
\item iron abundance - an enhanced Milky Way-like star formation creating an increased iron abundance.
\end{enumerate}

We do not believe the observed distribution differences are due to selection effects. The quasars observed with 
Hectospec were randomly selected from the photometric catalogues. 
The control sample was selected to match the redshift and magnitude distributions of LQG quasars.
However, there is a slight difference in the magnitude ranges, due to the magnitude limit of SDSS, 
shown in Fig. \ref{fig:W2400_mag}. Seven quasars
within the LQG are fainter than the control sample by $<$ 0.5 magnitudes.
However, the correlation between \textit{UV} Fe~\textsc{ii} and the quasar luminosity is still debated. 
Some studies have found an inverse Baldwin effect 
in the optical Fe~\textsc{ii} emission, with the EW Fe~\textsc{ii} emission 
increasing with the continuum emission \citep{Kovacevic2010, Dong2011, Han2011}. 
For the \textit{UV} Fe~\textsc{ii}, no significant correlation has been observed between 
the \textit{UV} Fe~\textsc{ii} and the quasar luminosity or 
$L/L_{Edd}$ \citep{Dong2011,Sameshima2009}.\footnote{The significance 
does increase, though still weak, if the \textit{UV} continuum is used to calculate the 
luminosity which is expected as the \textit{UV} Fe~\textsc{ii} 
is powered by the continuum at shorter wavelengths to the optical continuum.}

To investigate any effect of the magnitude on our data, a magnitude limit of $17.98 < g < 20.56$ 
was applied to both samples. The Mann-Whitney test gives a P = 0.0007 showing that even 
with a magnitude-limited sample which further limits the sample size, the distributions 
of the LQGs field and control samples are still different. 

The second possible explanation is an difference in dust properties between the LQG and the 
control sample causes the differences observed. As an excess of dust in the LQG region would 
reduce the UV emission, we do not believe this difference is due to dust emission. 
For dust emission to have an effect on our results, the control sample would have to see 
evidence of an steeper extinction law. However as the control sample consists of quasars from 13 different areas,  
it would require large scale special dust properties with the LQG field, which is unlikely. 
Since there is now a consensus that higher rates of star formation are seen in overdense environments at z $>$ 1
\citep[e.g.][]{Farrah2006, Amblard2011}, we think it very unlikely that ISM dust is the
cause of this difference, since if dust were causing the effect we'd expect the very high EW
systems to be found in the field.

The third and fourth options are Ly$\alpha$ fluorescence and microturbulence, which are 
additional mechanisms within the BLR believed to increase the Fe~\textsc{ii} emission. 
Again we do not believe this is the case as the control sample was selected to have similar quasar properties. 
As mentioned above, the small differences in magnitude are unlikely to be the cause of the 
distribution differences.

An increase in Ly$\alpha$ emission can cause an increase in the \textit{UV} 
Fe~\textsc{ii} emission \citep{Sigut2003,Sigut2004,Verner2004}. 
As the width of the Ly$\alpha$ increases, it
overlaps with numerous Fe~\textsc{ii} lines within the wavelength range 1212-1218 \AA.
These lines are excited, and when they decay produce emission in the \textit{UV}
Fe~\textsc{ii} region, 2200-2700 \AA. Increasing the Ly$\alpha$ emission will therefore
increase the \textit{UV} Fe~\textsc{ii} emission. In fact, \citet{Sigut1998} found that
Ly$\alpha$ fluorescent excitation can more than double the \textit{UV} Fe~\textsc{ii}
flux.

\begin{figure}
\includegraphics[width=60mm,angle=-90]{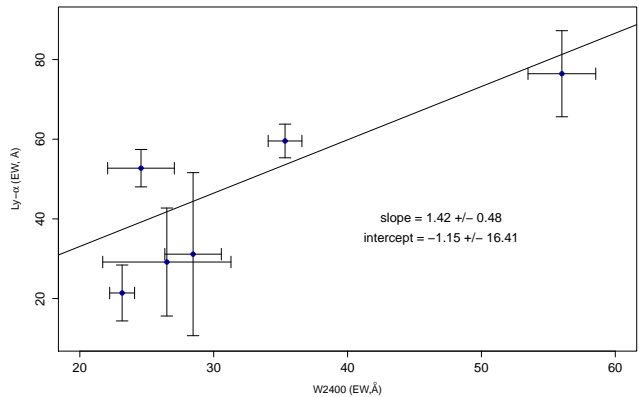}
\caption{\small{Comparing the equivalent widths for the Fe~\textsc{ii} emission from the
optical spectra and the Ly$\alpha$ emission from \textit{GALEX} \textit{UV} spectra. The
line fitted shows Ly$\alpha$\_EW = W2400\_EW, fit using least squares weighted
regression.}} 
\label{fig:LyA} 
\end{figure}

Low resolution $R\sim 90$  \textit{GALEX UV} spectra which cover the Ly$\alpha$ emission exist
for six of the quasars (program GI5-059, Williger et al.).
Fig. \ref{fig:LyA} shows the
correlation between the Fe~\textsc{ii} EW measurements and the equivalent widths of the
Ly$\alpha$ emission line. The line drawn is the weighted (using both sets of errors) 
least squares best-fit. The Pearson correlation coefficient between the Ly$\alpha$ and
the Fe~\textsc{ii} is 0.830 $\pm ~0.14$.
There is a suggestive trend for quasars with higher Ly$\alpha$ emission to have stronger
Fe~\textsc{ii} emission, as predicted \citep[e.g][]{Sigut2003,Sigut2004,Verner2004}. 
However, there are only six spectra here with \textit{GALEX}
Ly$\alpha$ emission. This fit is highly dependent on the presence of qso425 (which has the 
largest W2400 EW) and not robust.

Though the Ly$\alpha$ emission may influence the observed Fe~\textsc{ii} emission, there is no reason 
to believe the quasars within the LQG field have increased Ly$\alpha$ emission compared to randomly selected 
quasars. However, more data of quasar Ly$\alpha$ emission in various environments would be needed to fully investigate this.
Equally with an overdense environments, the effect of other quasars and nearby galaxies 
is negligible compared to the emission from the accretion disc of the quasar. 

The Fe~\textsc{ii} flux strength can also be increased by microturbulence around the AGN
\citep{Vestergaard2001,Sigut2003, Sigut2004, Verner2003, Verner2004, Bruhweiler2008}.
Microturbulence (non-thermal random motions within a cloud's line emitting region;
\citealt{Bottorff2000a,Bottorff2000b}) spreads the line absorption
coefficient over a larger wavelength range \citep{Bruhweiler2008}, broadens the
Ly$\alpha$ emission lines, and therefore increases the \textit{UV} Fe~\textsc{ii}
emission observed. 
Microturbulence is occurs within the BLR. Large scale dynamic effects due to the large scale
environment are unlikely to have an affect on the BLR 
without causing observable differences in the host galaxy, such star formation rates and 
luminosity, which is not seen here as our control was designed to match the field sample.

Although these factors have been shown to influence the \textit{UV} Fe~\textsc{ii} emission, 
modelling needs to be completed for quasars in environment over a range of densities 
to study how Ly$\alpha$ fluorescence and microturbulence can change with environment.

The final option is that the observed difference is due to the host galaxy and the quasars simply 
illuminate this difference. As previously noted, the dependence of metallicity with environment 
is still highly debated, with some 
studies showing a weak but significant trend for galaxies in higher density regions (such 
as groups or clusters) to have higher metallicities.
Therefore if, as we assume the quasars in LQGs trace the overdense regions, we 
would expect the host galaxies to have greater metallicities.

Galaxies with old stellar populations have been found to favour higher density environments 
at z $\sim$ 0 \citep[e.g.][]{Balogh2004,Blanton2005} and z $\sim$ 1 \citep[e.g.][]{Cooper2006}. 
\citet{Martini2013} found AGN have evolved more rapidly in higher density environments 
than the field population.
This suggests, at high redshifts, star formation may occur in high density environments 
\citep[e.g.][]{Cooper2008}.
If so, this will increase the metals available in the vicinity of these quasars. 
To produce the observed Fe~\textsc{ii} (assuming abundance 
is the main factor), the hosts would have gone through a period of enhanced star 
formation between $2 < z < 3$, assuming it takes between 0.3 Gyr and 1 Gyr \citep{Hamann1993} for the 
required number of SNeIa to occur to create significant amounts of iron. This is during the 
peak epoch of star formation \citep[e.g][]{Lilly1996,Madau1996, Sobral2013}.

There is no significant enhancement in Mg~\textsc{ii} in the LQG quasars compared to the 
control sample. This is consistant with a Milky Way-like star formation as opposed to a starburst. A 
increase in quiescent star formation in some of the galaxies within the LQGs 
would produce an increase in the iron abundance with respect to the Mg~\textsc{ii}.

Within an overdense region, there could also be additional metal enrichment of the quasars from 
supernovae occurring the inter-cluster medium and within nearby galaxies. 
Supernovae have been shown to efficiently enrich the IGM over Mpc scales
\citep[e.g.][]{Domainko2004,Adelberger2005}. These metals may then accrete the 
quasar, further enriching the quasar host.

\section{SUMMARY}

There is a increase in Fe~\textsc{ii} emission in a candidate overdense region, indicating there may 
be a build up of iron. 
It is consistant with an increase in star formation in overdense region at 
high redshift. This star formation must have occurred at 2 $<$ z $<$ 3 for iron to be observed 
in these quasars.
Additionally surrounding galaxies in this dense region will release metals into the IGM, 
which can fall onto the quasar, producing an observed metal increase.

This will make published LQGs interesting regions in which to study the evolution of metals 
in high density regions and at high redshifts.

\section{Acknowledgments}

KAH would like to acknowledge and thank the STFC, the University of Central Lancashire and the
Obs. de la C\^ote d'Azur for their funding, and hospitality. Also KAH would like thank
Roger Clowes and Luis Campusano for their communications

This research uses data from the Hectospec instrument on the MMT.

The authors acknowledge support from the NASA GALEX program GI5-059, grant NNX09AQ13G.

This research has used the SDSS DR7QSO catalogue \citep{Schneider2010}. Funding for the
SDSS and SDSS-II has been provided by the Alfred P. Sloan Foundation, the Participating
Institutions, the National Science Foundation, the U.S. Department of Energy, the National
Aeronautics and Space Administration, the Japanese Monbukagakusho, the Max Planck Society,
and the Higher Education Funding Council for England. The SDSS Web Site is
http://www.sdss.org/.

The SDSS is managed by the Astrophysical Research Consortium for the Participating
Institutions. The Participating Institutions are the American Museum of Natural History,
Astrophysical Institute Potsdam, University of Basel, University of Cambridge, Case
Western Reserve University, University of Chicago, Drexel University, Fermilab, the
Institute for Advanced Study, the Japan Participation Group, Johns Hopkins University, the
Joint Institute for Nuclear Astrophysics, the Kavli Institute for Particle Astrophysics
and Cosmology, the Korean Scientist Group, the Chinese Academy of Sciences (LAMOST), Los
Alamos National Laboratory, the Max-Planck-Institute for Astronomy (MPIA), the
Max-Planck-Institute for Astrophysics (MPA), New Mexico State University, Ohio State
University, University of Pittsburgh, University of Portsmouth, Princeton University, the
United States Naval Observatory, and the University of Washington.

\end{document}